# Sensor Networks Architecture for Vehicles and Pedestrians Traffic Control

Ovidiu Banias*, Daniel-Ioan Curiac* and Radu-Emil Precup*

* Automatics and Applied Informatics Department, "Politehnica" University Timisoara, Bd.V. Parvan Nr.2, 300223 Timisoara, Romania, Phone: (+40) 256-40-3251, Fax: (+40)-256-40-3214, E-Mail: ovidiu.banias@aut.upt.ro

*Abstract* - *In this paper we present a sensor network based architecture for urban traffic management, hierarchically structured on three layers: sensing, processing& aggregation and control. On proposed architecture we define traffic decongestion methods for vehicles and also for pedestrians. Finally, we presented a case study on how traffic control can be implemented in a concrete situation, based on the proposed architecture, pointing future directions of development.*

*Keywords:* *wireless sensor networks, traffic decongestion, aggregation, traffic control.*

## I. INTRODUCTION

Traffic control is a major problem nowadays due to the increasing number of cars and pedestrians, intelligent traffic solutions being researched for better fluidization of vehicles and people in crowded cities. Avoiding traffic jams is beneficial to environment and economy, but also raises the problem of increasing demand in vehicles, leading to a greater amount and a real problem of decongestion [1]. Intelligent Transport Systems are a must for increasing traffic safety and for offering road decongestion solutions [2].

Nowadays, for traffic decongestion expensive technologies are used (video cameras, inductive loops, pneumatic tubes for vehicle detection, complicated infrastructure) and very hard to maintain [3]. For lowering the decongestion costs the idea of using wireless sensor networks was taken in consideration, offering cheaper implementation and maintenance cost. The major importance of wireless sensor networks in traffic control is gave by their property of easy deployment and easy replacement without raising problems of traffic jams due to system unavailability. Unfortunately these solutions are still in the research stage, road traffic control using sensor networks implying carefully development both of network architecture, communication protocols and techniques of data processing.

This paper is organized as follows. Section 2 describes the proposed wireless sensor network architecture structured on three layers. In section 3 we present the sensing layer, describing the type of sensors needed and their properties. Section 4 explains the data aggregation and processing layer, gathering information from first layer and interpreting it for next layer. Section 5 describes the last layer of traffic lights control, working on predefined models and real time information received from previous layers. In section 6 we present a case study and we conclude in section 7, also pointing future work.

## II. SENSOR NETWORK ARCHITECTURE OVERVIEW

The sensor network architecture for road traffic control we propose in this paper is a general architecture, having wide applicability in urban road traffic management. We focus on vehicle traffic decongestion, not forgetting about pedestrians and other special cases. We present architecture for "simple crossroad traffic control structured on three layers in Fig. 1. The first layer is represented by the wireless sensing nodes, capturing vehicle and pedestrian movement and than sending information to the next layer.

The second layer is the "brain" of the wireless sensor network, receiving and aggregating information from the first layer, and then processing the data for the next layer. The third layer is represented by the traffic lights control, running implemented decongestion algorithms and changing lights according to the real time information received from the second layer. The architecture proposed will not contain motes deployed on the cars like presented in other work [4][5]. The main reason is that it is impossible to be sure the vehicles in the city have motes installed and also working. Supposing the whole vehicles in a city are equipped with motes, there will always be an important number from other cities and other countries without any mote. Another drawback of deploying motes on every vehicle, even by making it a standard and the deployment being done by the vehicles manufactures, is that it is impossible to control if those motes are working or not.

The third layer offer traffic decongestion solutions, changing traffic lights installed in the city crossroads. Road traffic control will depend on real time information from second layer and also on generated traffic statistics in different time intervals.

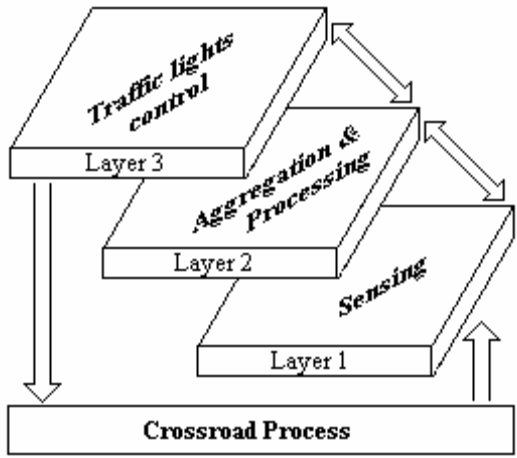

Fig. 1. Sensor network traffic control architecture layers

III. SENSING LAYER

Sensing layer is represented by a wireless sensor network, the motes being equipped with corresponding sensors for detecting and measuring traffic information. It is very important for the control system, the sensing layer to be wireless in order to achieve a low maintenance cost, just replacing the motes or maybe just changing batteries. Due to their wireless property, the motes could be deployed simple and in short time, without laborious cuts in the asphalt or pavement, avoiding traffic jams in crowded cities.

In Fig. 2 we present a simple crossroad sensor network architecture equipped with wireless sensors for detecting both vehicles and pedestrians.

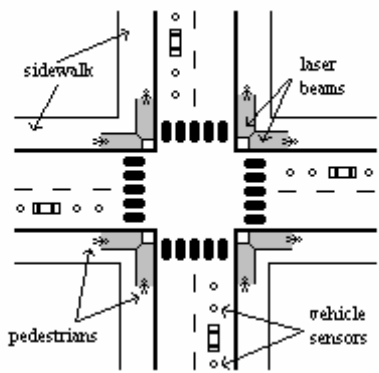

Fig. 2. Simple crossroad sensor network architecture

A. *Vehicles Sensing*

Vehicle detection with sensor networks is a new technology with already published results by Berkeley University Labs [6]. The most accurate sensors for vehicle detections are the magnetometers, detecting disturbances in earth magnetic fields in the presence of ferrous objects. The main advantage of these sensors over acoustic ones is the environment free property, vehicles acoustic signals being influenced by noises from other vehicles, and depending of the weather. Using two magnetometers placed at a short distance on the same lane, a sensor network is able to measure the approximately speed of the vehicles passing over them. The wireless sensor network should be able to detect special vehicles like ambulance, police, and fire-trucks. These vehicles should be able to transmit their presence by radio to the wireless networks deployed in crossroads, or should be equipped with special sensors detectable by the motes deployed on the road. With this kind of information the traffic control layer can take decisions to decongest traffic in the reported areas, setting "green light" for the special cars.

Vehicle sensors should be wireless, so used sensors could be chosen between magnetometers, acoustic sensors, and laser beams. The sensors should be placed on the road, along each lane approaching the crossroad. Depending on the traffic control strategy, only few sensors could be placed on one lane or sensors could be deployed in a large number beginning from a distance of hundreds of meters from the crossroad (repeaters should be installed to retransmit the information to the access point). In this last case, vehicle traffic is very attentive measured, the traffic control center receiving a large amount of information. The system should be able to immediately report traffic jam at lane level, due to broken vehicles, crashes or other reasons. The wireless sensor network will detect this problem by magnetometers, observing stationing vehicles for a period of time greater than the lane corresponding threshold.

B. *Pedestrians Sensing*

Another important problem in big cities is the detection of pedestrians. Mostly we need to decongest vehicle road traffic, but pedestrians must be taken in consideration as well, during traffic control. "Zebra" crossing is not a solution, because pedestrians passing it at an interval of couple of seconds could jam vehicle traffic. "Pelican" crossing (the pedestrians push a button to alert the system that they want to cross the street) is a good system, but also can be substituted with automatic sensing of pedestrians or maybe a combination between "push button" and automatic sensing. As people presence could not be detected by magnetometers, solutions used nowadays are detectors incorporated in the pavement [7]. The main disadvantage of these solutions, forgetting about maintenance problems, is that if sensing area is too big, than the system could not know if pedestrians are crossing street A or street B (see Fig. 3), and if sensing area is too small, than the system will be able to detect only few pedestrians. In the pedestrian sensing architecture we propose, the sensing areas colored in gray in Fig. 3, should be delimited from the rest of the sidewalk by small fences. With this delimitation, the system will be able to sense a bigger area of pedestrians waiting to cross the street, and not mistaken these pedestrians to the others just passing by the sidewalk. For example, sensing area A1 (in Fig. 3) will detect the number of pedestrians waiting and together with the pedestrians sensing area across the street will sum up the real time number of pedestrians willing to pass street A (see Fig. 3) . The delimited sensing areas could be seen as simple pedestrian lanes for street crossing waiting. In the proposed solution, based on small fences, pedestrian traffic will not be constrained, just structured for street crossing. To lower the cost and also to reach a better maintenance level, microware and laser beam sensors could be used in pedestrians sensing areas.

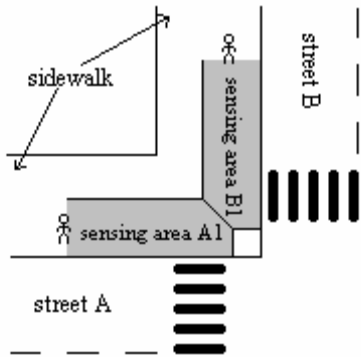

Fig. 3 Pedestrians sensing area

Sensing motes will transmit traffic information to the next layer represented by the access point, when vehicles are moving over them and also on demand by the access point. The transmition/reception will be done by radio using radio transceivers, both on motes and access point. PEDAMACS [8] (Power efficient and delay aware medium access protocol for sensor networks) is proven a good medium access protocol and could be used for transmitting traffic information to the access point. First, the access point creates a topology from the motes in the area and elaborates a time schedule for mote transmissions, and second, sends the schedule to each mote from the topology. Crossbow [9] is nowadays the market leader in wireless sensor networks components production. MICA2 or MICA2DOT could be a choice in wireless sensor network implementation because of their small size and proven reliability.

## IV. DATA PROCESSING AND AGGREGATION LAYER

The second layer is represented by the Access Point (AP). This is a special node of the sensor network, with capacity of data processing and data aggregation. Due to the need of computational power for data processing, and good transfer speed to the next layer, this node should be wired. The Access Point should be placed in each monitored crossroad, in order the motes to be able to communicate by radio with it (Fig. 4). Traffic information should be aggregated depending on traffic lights color, lane number, calculating the averages for different scenarios.

Data aggregation is essential for quick traffic lights control solutions. Instead of transmitting to the control layer every sensed information from deployed sensors, the Access Point will select and send only useful information. Redundant information will not be sent, possibly stored for statistics. Processed and aggregated information is sent after a well defined schedule to the next layer of traffic lights control. The second layer will send only requested information, avoiding any possible delays in traffic control generated by sending lot of packets from this layer to the next one. In Table 1 we present an example of the vehicle aggregation parameters per lane, on a defined test case. First of all, information is aggregated depending on traffic lights color (green/red). For each lane, the time passed from beginning of green light and time waiting from beginning of red light, are predefined values in seconds representing the interval of time from last measurement; when traffic color changes, these intervals will be truncated. For green light color, the number of cars that passed and for red light, the number of waiting cars added to the queue, are important for monitoring traffic flow and produce events for traffic control layer. Queue length between certain time intervals is another important measuring method of traffic status, giving information about congested lanes at certain crossroad for real time traffic control. Average speed is measured easily by two neighbor motes placed on the road, being a good parameter for traffic flow statistics and control. The number of pedestrians crossing and waiting should also be aggregated per time intervals for a better real time traffic control. For better usage, the number of pedestrians waiting to cross one street will be summed up from the number of pedestrians waiting on both sides of the street. We present in section VI a case study explaining the aggregated data.

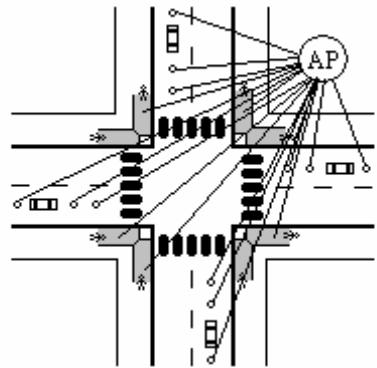

Fig. 4. Crossroad Access Point

## V. TRAFFIC LIGHTS CONTROL LAYER

This layer is meant to offer traffic decongestion solutions, changing traffic lights installed in the city crossroads. Road traffic control will depend on real time information from second layer and also on generated traffic statistics in different time intervals. We propose implementing the control layer as real time knowledge based expert system. In Fig. 5 we present the Traffic Lights Control Layer architecture.

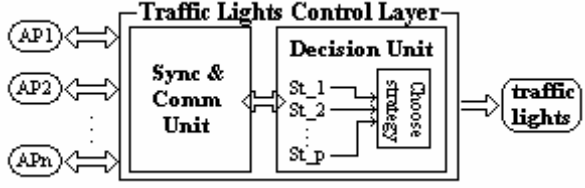

Fig. 5. Traffic Lights Control Layer

The Traffic Lights Control Layer is composed from two main blocks: Syncronization & Communication Unit (SCU) and Decision Unit (DU). SCU transmits and receives requested information from the access points and also commands the synchronization of the access points and wireless sensor network. The needed information from sensing network is requested through parameters, filtering only the important data from aggregation points. The Decision Unit is the brain of the traffic control layer, and based on statistics and real time information received from the SCU, decisions are made about what traffic control strategy should be chosen. Control strategies are chosen by real time priorities for vehicles and also for pedestrians.

An example of Decision Unit pseudocode is presented as follows.

```
REPEAT
  IF timer_AccPointReading THEN
    CALL ReceiveFromAP(parameters)
  END IF
  IF timer_ChagedLights OR ReceivedFromAP THEN
    CALL ControlSolution
  END IF
UNTIL FALSE
SUB ReceiveFromAP(parameters)
  // receive readings from Access Point - only requested parameters
  ...
END SUB
SUB ControlSolution // Classifier
  //ChoseStrategy
  UpdateFitnessFunction
  CreatePriorities
  Select Max(Priorities)
  Select Strategy
END SUB
```

Accepting the following nomenclature:

$IN_i$ - street number $i$ entering the crossroad,

$OUT_i$ - street number $i$ leaving the crossroad,

$IN_i OUT_j$ - traffic event pointing out that the vehicle enters the crossroad from street number $i$ and leaves the crossroad on street number $j$,

$C_i$ - traffic event highlighting that the pedestrians are crossing the street number $i$,

we define the following set of possible events in the crossroad, $E$, useful for later defined events graph:

$$E = \{IN_i OUT_j / i = \overline{1, nrIN}, j = \overline{1, nrOUT}\} \cup$$
$$\cup \{C_j / j = \overline{1, nrc}\} \qquad (1)$$

where:

$nrIN$ - number of streets entering the given crossroad,
$nrOUT$ - number of streets leaving the given crossroad,
$nrc$ - number of pedestrian crossings in given crossroad.

We will define also the graph $G(E, A)$ of possible events in one crossroad. In this context, in relation with (1) $E$ is the set vertices, each vertex being represented by one event, and $A$ is the adjacency matrix, defining edges between vertices:

$$A_{x,y} = \begin{cases} 1, E_x \& E_y \text{ are not criss-crossed} \\ 0, E_x \& E_y \quad \text{are criss-crossed} \end{cases}, \qquad (2)$$

with $A_{x,y}$ - elements of the adjacency matrix of graph $G$. As shown in (2), there is an edge between two events if those events are criss-crossing each other. From the traffic control point of view, one edge between two events means those events should always have different traffic lights color.

We will define as follows the fitness function for each event. According to the value returned, the control layer will chose witch events will get "green" and witch will wait for the next selection. The importance of this function is that events could be prioritized depending on parameters choice / set, and the fitness function $F_{Ex}$ is defined in (3):

$$F_{Ex} : R^4 \to R, \ F_{Ex}(p_1, p_2, p_3, p_4) = x, \qquad (3)$$

where: $p_1$, $p_2$, $p_3$ and $p_4$ are function variables and $x$ is the fitness index of event $E_x$. This function may be used as cost function in an optimization problem with the variables $p_1$ (line length only for vehicles), $p_2$ (number of seconds waiting on red), $p_3$ (number of seconds passing on previous green), $p_4$ (number of vehicles/pedestrians waiting) and appropriately defined constraints.

For each crossroad we define a changing traffic lights counter for each possible event, in case of event number $i$ from set $E$ the counter being referred to as $T_i$. $T_i$ is managed by the access point and incremented each time the traffic lights charge.

As a general traffic rule we state that no event should wait more than a defined threshold and no event should wait until all other events were raised at least two times, excepting the case that event is not requested (no cars or no pedestrians waiting depending on event type). When traffic lights change, $T$ counters are important for selecting at least one event than was not raised lately. The traffic control layer will chose at least one event with the property (4):

$$T_i = \min(T). \qquad (4)$$

Choosing all events with minimum counters value, may raise criss-crossing problems, so from the set of events with the property described above, should be selected only the ones not criss-crossing each other. Another important observation is that after selecting events with minimum counter values, there could be other events with bigger counter values that could be raised together with selected ones because the not criss-crossing rule is respected.

Traffic control and decongestion should be made according with defined events priorities parameters like: fitness function $F$, events graph $G$ and counters $T$. When traffic lights colors need to be changed, according to priorities defined above, from the set of events $E$, every subset $E'$ of events could get "green" as long as $E'$ is the set of vertices of subgraph type $H$ with corresponding properties. This is formalized in (5):

$$H(E', A') \subset G(E, A) / E' \subset E, A' \equiv O. \qquad (5)$$

If given crossroad needs a vehicle traffic decongestion then the subset $E'$ should be chosen only among $IN_i OUT_j$ type events. In the other case of pedestrians decongestion the subset $E'$ should be chosen only among $C_i$ type events, with the possibility of setting "green" to all pedestrian events in given crossroad. In this special case of pedestrian decongestion, letting pedestrians crossing the whole crossroad by passing even through the middle of it could be a solution [10]. The system should inform drivers and pedestrians about the changing lights with at least 2-3 seconds in advance with countdown timers. The amount of seconds between countdown timers start and the lights change should not be too small (traffic attendants should have time to prepare) and not too big (real time traffic information could lead to changing control solutions as soon as possible). A good traffic decongestion method is to maximize the passing through intersection average speed by choosing longer red/green timers, to let the cars to increase the average speed per intersection. Shorter red/green timers imply more stops, traffic becoming fragmented, because of the lower average speed.

## VI. CASE STUDY

This section is dedicated to presenting a case study on a certain crossroad like described in Fig. 2. It is a simple crossroad entering four streets and leaving other four, and pedestrian crossings are also in number of four. Like presented in section III, the crossroad is equipped with wireless sensor node for pedestrian and vehicle sensing. In our example we consider vehicle sensors deployed along 200 m from the crossroad, with the distance of 4 meters between them. The access point will have the role of synchronizing sensor node by Pedamacs[7] schedule, receive sensed information, aggregating it, and then sending it to the control layer. The real time aggregated information is essential for quick traffic control, and Table I presents an example of traffic measurement on one lane between intervals of 5 seconds, regarding number of cars, average speed and queue length. From the first three rows of the table can be noticed the increasing average speed for vehicles passing on green color and also the shrinkage of queue length. The last three rows of the table shows the aggregated information for the red color, the number of cars added to the queue and the queue length between intervals of 5 seconds. The queue length being very important in traffic decongestion, informs the system about exceeding certain queue length threshold.

For the crossroad presented in Fig. 2, we note the four streets entering the crossroad from $IN_1$ to $IN_4$ in clockwise order, and streets leaving the crossroad from $OUT_1$ to $OUT_4$, by the rule "each $OUT_i$ near $IN_i$". The obtained set of events is presented in (6), and the corresponding adjacency matrix of graph $G$, in Table II. A $T$ counters example is presented in Table III, with a distance between counters values lesser than the number of elements in set E. The minimum counters value is 6, so each of the events could be chosen along with the corresponding subgraph $H$ (5) for getting green color. In the end, the traffic control strategy is chosen depending the event priorities and chosen fitness function.

*TABLE I. Vehicle traffic aggregation parameters per lane.*

|  | green light | | | red light | |
|---|---|---|---|---|---|
| Time pattern [hh:mm:ss.s] | No. of cars in $\Delta tp$ | Average speed [m/s] | Queue length [m] | No. of cars in $\Delta tw$ | Queue length [m] |
| 12:23:52.55 | 19 | 7 | 200 | | |
| 12:23:57.76 | 26 | 10 | 127 | | |
| 12:23:59.80 | 12 | 13 | 30 | | |
| 12:24:05.05 | | | | 15 | 60 |
| 12:24:10.07 | | | | 13 | 121 |
| 12:24:15.01 | | | | 7 | 160 |

From (1) the set of events of presented crossroad will have 16 elements, 12 for vehicles and 4 for pedestrians according to (6):

$$\begin{aligned} E = \{ &IN_1 OUT_2, IN_1 OUT_3, IN_1 OUT_4, \\ &IN_2 OUT_3, IN_2 OUT_4, IN_2 OUT_1, IN_3 OUT_1, \\ &IN_3 OUT_2, IN_4 OUT_1, IN_4 OUT_2, IN_4 OUT_3, \\ &C_1, C_2, C_3, C_4 \} \end{aligned} \qquad (6)$$

The steps presented above are repeated before each changing traffic lights timer is fired, selecting one of the most adequate traffic decongestion strategy for given crossroad. In the end, the traffic control strategy is chosen depending the event priorities and according to selected fitness function.

## VII. CONCLUSIONS

Urban traffic congestion has apparently developed into an untouchable issue. The most important contributing factors to decongestion are the innovative and efficient solutions that will make a real difference.

TABLE II. Adjacency matrix of graph G.

|     | E1 | E2 | E3 | E4 | E5 | E6 | E7 | E8 | E9 | E10 | E11 | E12 | E13 | E14 | E15 | E16 |
|-----|----|----|----|----|----|----|----|----|----|-----|-----|-----|-----|-----|-----|-----|
| E1  |    |    |    | 1  | 1  |    |    | 1  | 1  | 1   | 1   |     | 1   | 1   |     |     |
| E2  |    |    |    | 1  | 1  |    | 1  |    |    | 1   | 1   | 1   | 1   |     | 1   |     |
| E3  |    |    |    | 1  |    | 1  |    |    |    |     |     |     | 1   |     |     | 1   |
| E4  | 1  | 1  |    |    |    |    | 1  | 1  |    |     | 1   | 1   |     | 1   | 1   |     |
| E5  | 1  | 1  | 1  |    |    |    | 1  | 1  |    | 1   |     |     |     | 1   |     | 1   |
| E6  |    |    |    |    |    |    | 1  |    | 1  |     |     |     | 1   | 1   |     |     |
| E7  |    | 1  | 1  | 1  | 1  |    |    |    |    | 1   | 1   |     |     |     | 1   | 1   |
| E8  | 1  |    |    | 1  | 1  | 1  |    |    |    | 1   |     |     | 1   |     | 1   |     |
| E9  | 1  |    |    |    |    |    |    |    |    |     | 1   |     |     | 1   | 1   |     |
| E10 | 1  | 1  |    |    | 1  | 1  | 1  | 1  |    |     |     |     | 1   |     |     | 1   |
| E11 | 1  | 1  |    | 1  |    |    | 1  | 1  | 1  |     |     |     | 1   |     |     | 1   |
| E12 |    | 1  |    | 1  |    |    |    |    |    |     |     |     |     |     | 1   | 1   |
| E13 | 1  | 1  | 1  |    |    | 1  |    | 1  |    | 1   |     |     |     |     |     |     |
| E14 | 1  |    |    | 1  | 1  | 1  |    |    | 1  |     | 1   |     |     |     |     |     |
| E15 |    | 1  |    | 1  |    |    | 1  | 1  | 1  |     |     | 1   |     |     |     |     |
| E16 |    |    | 1  |    | 1  |    | 1  |    |    | 1   | 1   | 1   |     |     |     |     |

TABLE III. T counters.

| T1 | T2 | T3 | T4 | T5 | T6 | T7 | T8 | T9 | T10 | T11 | T12 | T13 | T14 | T15 | T16 |
|----|----|----|----|----|----|----|----|----|-----|-----|-----|-----|-----|-----|-----|
| 21 | 20 | 17 | 16 | 16 | 21 | 18 | 19 | 16 | 22  | 17  | 16  | 18  | 16  | 17  | 20  |

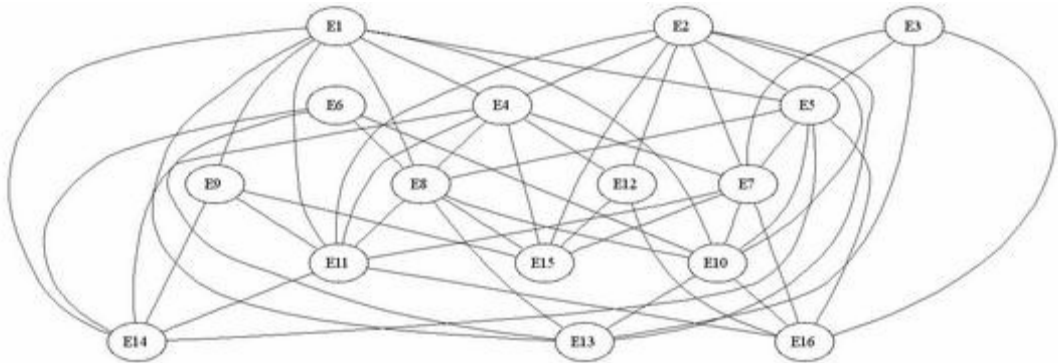

Fig. 6. Graph *G(E,A)*

The sensor network based architecture presented in this paper considers three factors that can improve the situation: wireless sensing, processing and aggregating modules, real time adaptive signal control.

Future work will be developed in the direction of researching adequate fitness function formula for static crossroads, with later applications to synchronized crossroads in whole city.